\documentclass[%
 prl,superscriptaddress,
reprint,
 amsmath,amssymb,
 aps,
]{revtex4-2}

\usepackage[resetlabels]{multibib}
\newcites{SI}{null}

\usepackage{graphicx}
\usepackage{dcolumn}
\usepackage{bm}
\usepackage{hyperref}
\usepackage{siunitx}

\usepackage{bbm}
\usepackage{ulem}
\usepackage{color}
\definecolor{mygreen}{rgb}{0,0.5,0}
\definecolor{mydkgreen}{rgb}{0,0.35,0}
\definecolor{mygrey}{rgb}{0.5,0.5,0.5}
\definecolor{myred}{rgb}{0.75,0,0}
\definecolor{myblue}{rgb}{0,0,0.75}
\definecolor{mymagenta}{cmyk}{0,1,0,0.12}
\definecolor{mycyan}{cmyk}{1,0,0,0.12}
\definecolor{myorange}{rgb}{1,0.5,0}
\definecolor{myviolet}{rgb}{0.5,0.0,0.75}
\definecolor{mybrown}{rgb}{0.75,0.5,.5}

\newcommand{\cX}{{\cal X}}

\renewcommand{\and}{a_n^\dagger}

\newcommand{\SNR}{{\rm SNR}}
\renewcommand{\SNR}{\zeta}

\newcommand{\SQMZI}{{\rm SQ-MZI}}

\newcommand{\phisig}{\varphi_{\rm sig}}
\newcommand{\phisighat}{\hat{\varphi}_{\rm sig}}

\newcommand{\MZIsub}{{\rm M}}

\begin{document}

\preprint{APS/123-QED}

\newcommand{\thetitle}{SU(2)-in-SU(1,1) Nested Interferometer for Highly Sensitive, Loss-Tolerant Quantum Metrology}
\title{\thetitle}

\newcommand{\myaffiliation}{\affiliation}
\newcommand{\ECNU}
{
	\myaffiliation{Quantum Institute of Light and Atoms,Department of Physics, East China Normal University, Shanghai, 200241, People's Republic of China}}
\newcommand{\SJTU}
{
	\myaffiliation{Department of Physics and Astronomy, Tsung-Dao Lee Institute, Shanghai Jiao Tong University, Shanghai 200240, People's Republic of China}}

\newcommand{\ICFO}
{
	\myaffiliation{ICFO-Institut de Ci\`encies Fot\`oniques, The Barcelona Institute of Science and Technology, 08860 Castelldefels (Barcelona), Spain}}
\newcommand{\ICREA}
{
	\myaffiliation{ICREA -- Instituci\'o Catalana de Recerca i Estudis Avan\c{c}ats, 08010 Barcelona, Spain}
}

\newcommand{\HDU}
{
	\myaffiliation{Department of Physics, Hangzhou Dianzi University, Hangzhou 310018, China}
}

\newcommand{\UC}
{
	\myaffiliation{Department of Physics, Niels Bohr Institute, University of Copenhagen, DK-2100 Copenhagen, Denmark.}
}

\newcommand{\SUTC}
{
	\myaffiliation{Shenzhen Institute for Quantum Science and Engineering and Department of Physics, Southern University of Science and Technology, Shenzhen 518055, China}
}

\newcommand{\PDU}
{
	\myaffiliation{Department of Physics, City University of Hong Kong, 83 Tat Chee Avenue Kowloon, Hong Kong, People's Republic of China}
}

\newcommand{\CIC}
{
	\myaffiliation{Collaborative Innovation Center of Extreme Optics, Shanxi University, Shanxi 030006, People's Republic of China}
}

\newcommand{\SHR}
{
	\myaffiliation{Shanghai Research Center for Quantum Sciences, Shanghai 201315, People's Republic of China}
}

\author{Wei Du}
\SJTU
\ICFO

\author{Jia Kong}
\HDU
\ICFO

\author{Guzhi Bao}
\SJTU

\author{Peiyu Yang}
\SJTU

\author{Jun Jia}
\UC

\author{Sheng Ming}
\SJTU

\author{Chun-Hua Yuan}
\ECNU

\author{J. F. Chen}
\SUTC

\author{Z. Y. Ou}
\PDU

\author{Morgan W. Mitchell}
\email[Corresponding author: ]{morgan.mitchell@icfo.eu}
\ICFO
\ICREA

\author{Weiping Zhang}
\email[Corresponding author: ]{wpz@sjtu.edu.cn}
\SJTU
\CIC
\SHR

\date{\today}

\newcommand{\Suppl}{Supplementary information}
\newcommand{\acronym}{\rm SISNI}
\newcommand{\boldacronym}{\textbf{SISNI}}

\begin{abstract}
{
We present experimental and theoretical results on a new interferometer topology that nests a SU(2) interferometer, e.g., a Mach-Zehnder or Michelson interferometer, inside a SU(1,1) interferometer, i.e., a Mach-Zehnder interferometer with parametric amplifiers in place of beam splitters.  
This SU(2)-in-SU(1,1) nested interferometer (\acronym)} simultaneously achieves high signal-to-noise ratio (SNR), sensitivity beyond the standard quantum limit (SQL) and tolerance to photon losses external to the interferometer, e.g., in detectors. We implement a SISNI using parametric amplification by four-wave mixing (FWM) in Rb vapor and a laser-fed Mach-Zehnder SU(2) interferometer. We observe path-length sensitivity with SNR \SI{2.2}{\decibel}  beyond the SQL at power levels (and thus SNR) 2 orders of magnitude beyond those of previous loss-tolerant interferometers.  We find experimentally the optimal FWM gains and find agreement with a minimal quantum noise model for the FWM process. The results suggest ways to boost the in-practice sensitivity of  {high-power interferometers, e.g.,} gravitational wave interferometers, and may enable high-sensitivity, quantum-enhanced interferometry at wavelengths for which efficient detectors are not available. 
\end{abstract}
\pacs{Valid PACS appear here}
\maketitle


The use of squeezing \cite{CavesPRD1981,SlusherPRL1985} and entanglement \cite{HollandPRL1993, MitchellN2004} allows advanced interferometers to detect signals that would otherwise be buried in quantum mechanical noise. The signal-to-noise ratio (SNR) is a central figure of merit in any such sensing application. The signal strength can be increased by using a larger flux of photons, while quantum noise can be reduced below the shot-noise level using nonclassical states of light \cite{CavesPRD1981}.  Approaches that use single squeezed beams and SU(2) interferometers, e.g. Michelson or Mach-Zehnder interferometers \cite{yurke19862}, have been successful in producing more than \SI{10}{dB} of noise suppression  \cite{mehmet2011squeezed, vahlbruch2016detection}, and have been applied in interferometers with very high photon flux  \cite{LIGONP2011, AasiNP2013Short}, to achieve both large and quantum-enhanced signal-to-noise ratios. In gravitational wave detectors, a variety of effects limit the useful power  such that both high flux and nonclassical input are required to achieve maximum sensitivity  \cite{TsePRL2019Short}.

\begin{figure*}[t]
\centering
\includegraphics[width=1 \textwidth]{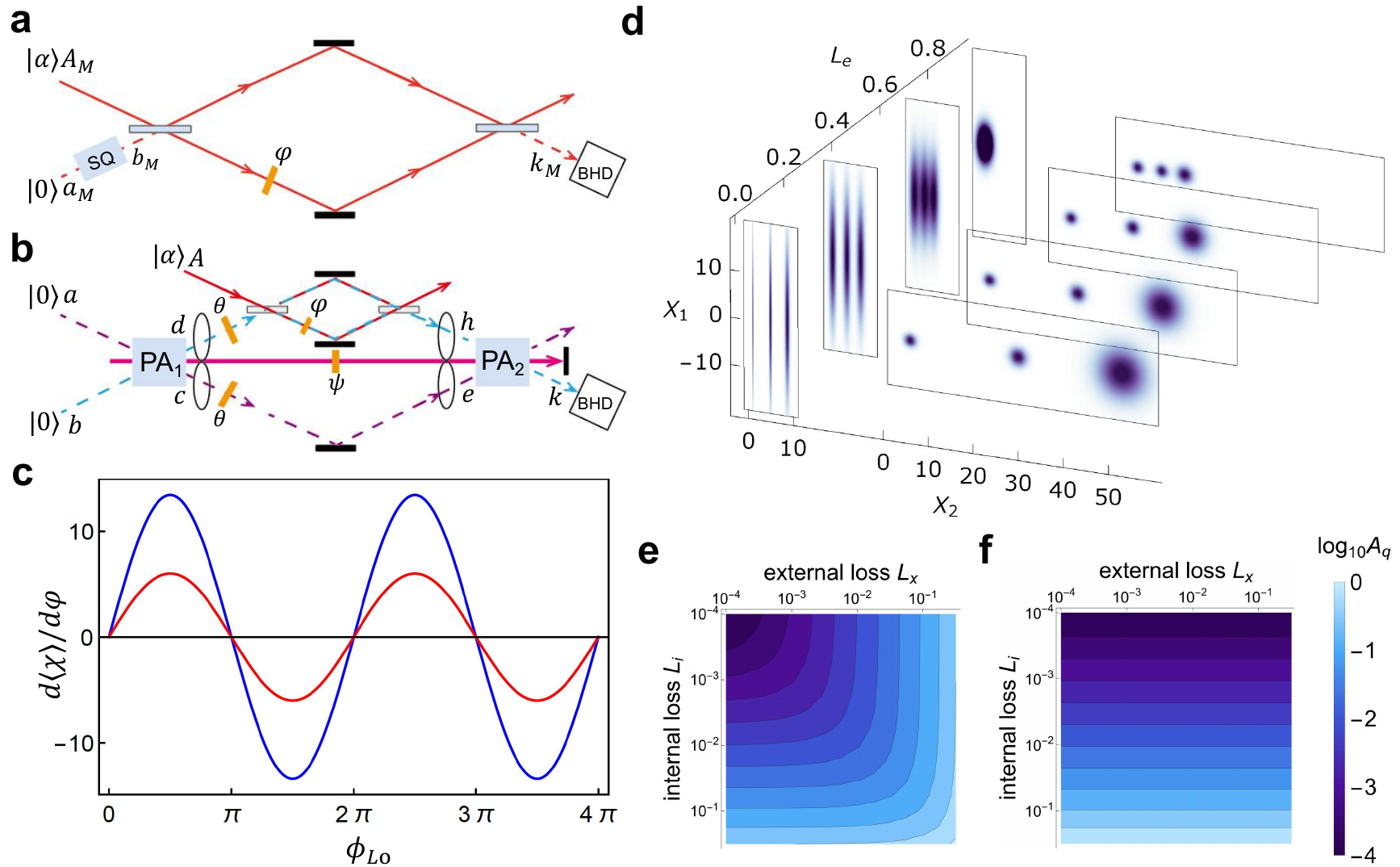} 
\caption{Comparison of \SQMZI~and \acronym~{in theoretical simulations}. (a), (b) Topology of \SQMZI~and \acronym, respectively. BHD, balanced homodyne detection.  Insets show the various input states in the amplitude quadrature ($X_1$) phase quadrature ($X_2$) phase space. For both interferometer types, a coherent state $|\alpha \rangle$ feeds the ``bright input'' port of the SU(2) interferometer, while a nonclassical state feeds the ``dark input'' port. (c) Signal gain $d\cX/d\phi$ versus local oscillator phase $\phi_\mathrm{LO}$ for \SQMZI~(red) and \acronym~(blue), for the lossless case. Both configurations have $|\alpha|^2 = 36$ (d) {Shading indicates Wigner}  distributions for the output mode {$\hat{k}$ and $\hat{k}_{M}$} versus signal phase $\varphi$ (three equispaced $\varphi$ values are shown in $X_1$, $X_2$ phase space) and versus external loss $L_{e}$.  For larger loss values, degradation of SNR, i.e., overlap of uncertainty areas, is evident in the \SQMZI~case, but not that of \acronym. (e), (f) Quantum sensitivity advantage $A_q \equiv \langle \delta \varphi^2 \rangle / \langle \delta \varphi^2 \rangle_{\rm SQL}$ in decibels, calculated from Eqs.~(\ref{SQ-MZI1}) and (\ref{SISI}),as a function of internal and external loss for (left) \acronym~ and  (right) \SQMZI. 
}
\label{fig:setup}
\end{figure*}

A current limitation of this approach is the loss of squeezing that accompanies optical losses, which break the photon-photon correlations of a strongly squeezed state.  In Aasi \textit{et al.} \cite{AasiNP2013Short}, for example, \SI{2.2}{dB} of extra sensitivity was achieved even though \SI{10.3}{dB} of squeezing was available, due in large part to the \SI{44}{\percent} system efficiency. Losses external to the interferometer
itself, including photodetection inefficiency, significantly impact the achievable quantum advantage \cite{AasiNP2013Short,TsePRL2019Short,AcernesePRL2019Short}.

An alternative approach is to use in-principle-noiseless quantum optical methods to amplify the interferometer signal, rather than aiming to suppress quantum noise. For example, the SU(1,1) interferometer \cite{YurkePRA1986} (SUI) 
has the topology of the Mach-Zehnder interferometer (MZI), but the splitting and recombination elements are parametric amplifiers (PAs), rather than passive beam splitters  \cite{hudelist2014quantum}. The SNR of the SUI can be insensitive to external loss, i.e., to losses outside of the loop formed by the two paths and PAs \cite{FrascellaNPJQI2021}. Similar advantages can be obtained in a ``truncated'' SUI that replaces the second PA with correlated phase-sensitive detection \cite{AndersonO2017}.


The SUI approach has its own limitations.  The PAs, as their name suggests, are {used as} amplifiers rather than light sources \textit{per se}.  To increase photon flux and signal strength, seed light is introduced into the upstream PA to stimulate the generation of bright two-mode squeezed beams.  While the PA process can be noiseless in theory, the best implementations to date are based on atomic four-wave mixing (FWM), which inevitably introduces additional noise \cite{gaeta1992quantum, schirmer1997quantum, boyd1994quantum,KauranenOC1993,IruvantiOC1995, davis1995excess, hsu2006effect}. This FWM noise grows faster with seed power than does the signal, thereby setting an intrinsic limit to SNR. This drawback has to date limited the phase-sensing light power to tens of microwatts in SUI interferometers  \cite{hudelist2014quantum, manceau2017detection, Anderson:17}. Similarly, nonlinear-optical effects limit the practical twin-beam power of FWM in optical fiber \cite{Liu:18}.

Here we propose and demonstrate a new SUI-based approach, the SU(2)-in-SU(1,1) nested interferometer (\acronym),  which combines advantages of SU(1,1) and SU(2) interferometry.  As illustrated in Fig.~\ref{fig:setup}(b), a SU(2) interferometer is nested inside a SU(1,1) interferometer: the ``signal'' beam from the upstream PA {(PA$ _{1} $)} is fed into the {dark input} port of a SU(2) interferometer, while a bright coherent state is fed into the \textit{bright input} port. The light emerging from the SU(2) {dark output} port is then recombined with the idler beam in the downstream PA (PA$ _{2} $).   
{The advantage of the SISNI can be explained as follows: in the \textit{dark fringe} condition, the SU(2) dark output is a noiseless copy of the dark input, displaced by an amplitude $\propto |\alpha| \varphi$. The displacement indicates the phase $\varphi$, boosted by the strong coherent state magnitude $|\alpha|$. The SU(1,1) interferometer, meanwhile, is a loss-tolerant detector of translations \cite{FrascellaNPJQI2021}. In this way, the SISNI achieves the large signal strength of SU(2) interferometry and the loss-tolerant quantum noise reduction of the SUI approach.
}


 In principle, the same advantages can be achieved with single-mode squeezers at the input (for sub-shot-noise operation) and at the output (for loss tolerance)  \cite{CavesPRD1981}. The SU(1,1) provides both functions in a simple implementation, {and also gives one the possibility to sense with one wavelength and detect at another \cite{Zeilinger1}.}  Another modified SUI was recently proposed, the so-called pumped-up SUI  
 \cite{szigeti2017pumped}, which achieves large photon flux by detecting interference of the signal, idler and also pump. This approach appears attractive for atomic interferometry, with possible implementation in spinor Bose-Einstein Condensates \cite{LinnemannPRL2016}  or hybrid atom-light systems\cite{PhysRevLett.115.043602}.

\begin{figure*}[tbp]
	\centering
	\includegraphics[width=18cm]{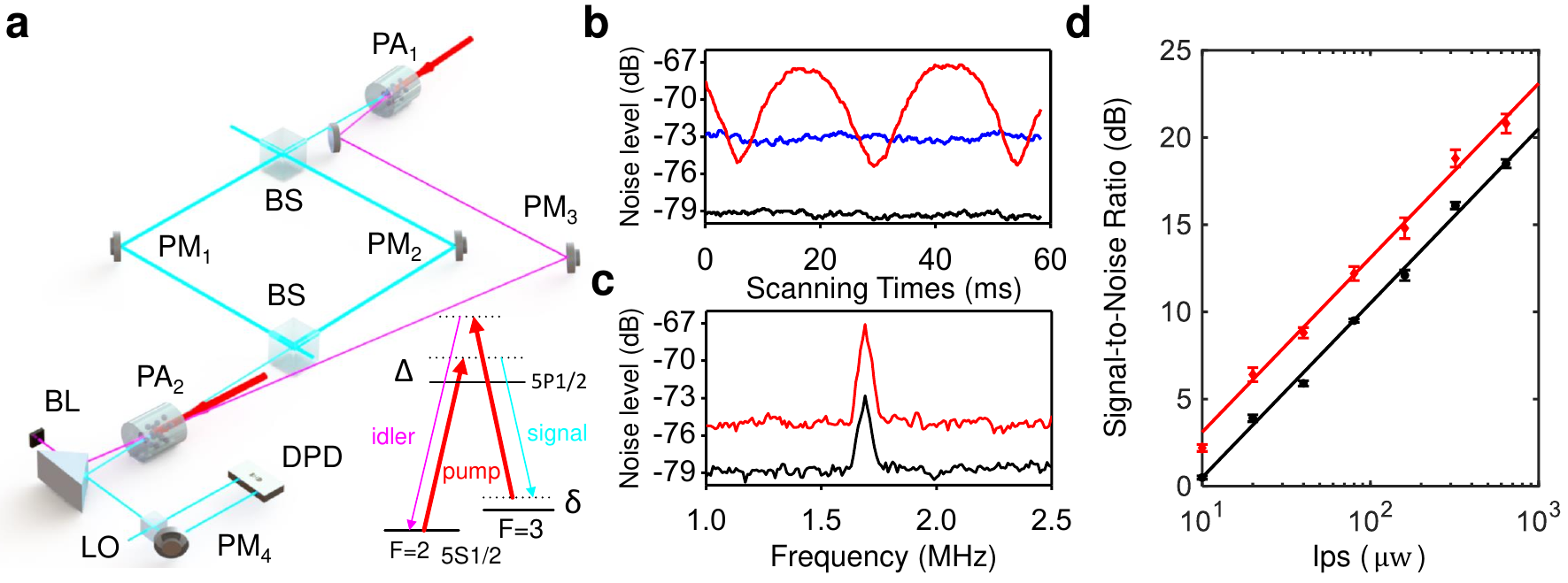}
	\caption{ Experimental demonstration of quantum enhancement of the \boldacronym. (a) Schematic of experiment. BS, 50/50 beam splitter; PM, mirror mounted with piezoelectric transducer as phase modulator; LO, local oscillator; DPD, differential photodetector; BL, beam block. Inset shows the double-$\Lambda$ level structure of the PA process, which uses the D1 line of $^{85}$Rb. The two pump beams (red arrows) are frequency degenerate, $\Delta = \SI{1}{\giga\hertz}$ and $\delta = \SI{2}{\mega\hertz}$. PA$_{1}$ and PA$_{2}$ act as a source of two-mode squeezing and amplifier, respectively. (b) Noise of the MZI and \acronym. Graph shows measured noise of the output phase quadrature in a bandwidth of \SI{100}{\kilo\hertz} about \SI{1.7}{\mega\hertz}, versus time as $\phi$, the relative phase of the PAs, is scanned using PM$_3$. MZI phase is locked to dark fringe. Traces show MZI SQL (black), implemented by blocking both PA pumps, MZI output amplified by PA$_2$ (blue), implemented by blocking only PA$_1$ pump, and \acronym~when MZI locked at dark fringe as the SU(1,1) phase is scanned (red).  (c) Measured noise spectra of MZI (black) and \acronym~(red) outputs, acquired as in (b), but with PAs relative phase locked to noise minimum and a sinusoidal phase signal at \SI{1.7}{\mega\hertz} applied via PM$_1$. (d)  Measured SNR, defined as spectral peak over white background level from spectra as in (c), versus phase-sensing light intensity ($I_\mathrm{ps}$) for MZI (black) and \acronym~(red).  Error bars show $\pm 1\sigma$ statistical variation. Lines show fits with ${\rm SNR} = A |\alpha|^2$. The QNG of PA$ _{1} $ and PA$ _{2} $ are G$ _{q1} = 4$ and G$ _{q2} = \SI{6}{\decibel}$, respectively.}
	\label{fig:Noisespectral}
\end{figure*}


The quantum optical performance of each strategy can be analyzed by considering cascaded linear input-output relations \cite{PRLSIRef, DuAPL2020}. For a  {SU(2) interferometer, e.g.} MZI, the input and output have the relation
\begin{eqnarray}\label{MZI-inout}
		\hat{k}_{\MZIsub}&=&\hat{N}_{\MZIsub}+\sqrt{\eta}\left[\cos(\frac{\varphi}{2})\hat{b}_{\MZIsub}-i\sin(\frac{\varphi}{2})\hat{A}_{\MZIsub}\right],
\end{eqnarray}	
where $\hat{N}_{\MZIsub}=\sqrt{L_{e}}\hat{D}_{\MZIsub}+\sqrt{L_{i}(1-L_{e})/2}(\hat{B}_{\MZIsub}-\hat{C}_{\MZIsub}) $ and $\eta=(1-L_{i})(1-L_{e}) $.  {A strong coherent state $| \alpha \rangle$ is injected into the MZI bright input  $\hat{A}_{\MZIsub}$, vacuum enters by modes $\hat{B}_{\MZIsub}$, $\hat{C}_{\MZIsub}$, and $\hat{D}_{\MZIsub}$,} $  L_{i} $ and $ L_{e} $ indicate {internal (affecting both beams of MZI) and external (after the second MZI beam-splitter) photon loss probabilities,} respectively. The dark fringe condition is $ \varphi = \varphi_0+\phisig $, where $\varphi_0 = 0$ is the {operating} point and $ \phisig\ll1$ is the signal phase.  {The term in $\sin(\phisig/2)\hat{A}$ contributes  the signal, which can be large if mode $\hat{A}$ contains a bright coherent state. In low-loss conditions, the term in $\cos(\phisig/2)\hat{b}_M$ contributes nearly all the noise; mode $\hat{A}$, being a coherent state, contributes unit noise scaled down by the small prefactor $\sin(\phisig/2)$.}

{A squeezed-light MZI (SQ-MZI), injects squeezed light into the MZI ``dark input,'' as shown in Fig.~\ref{fig:setup}(a).  This places} the mode $ \hat{b}_{\MZIsub} $ in Eq.~(\ref{MZI-inout}) {in} a vacuum squeezed state: that is, $\hat{b}_{\MZIsub}= G\hat{a}_{\MZIsub}+g\hat{a}_{\MZIsub}^{\dagger}$, where $G = \sqrt{1+g^2}$ is the amplification gain of the single-mode squeezer and $ \hat{a}_{\MZIsub} $ is a vacuum state. {This suppresses noise in one quadrature of mode $\hat{b}_{\MZIsub}$ and increases} the SNR. The SNR, after optimization of the squeezing and interferometer phases, is
{
\begin{eqnarray}
\label{SQ-MZI1}
\SNR_{\SQMZI} &=&\frac{\eta\left| \alpha\right| ^2 \phisig^2  }{1+\eta\left[ (G-g)^2-1\right]}.
\end{eqnarray} 
}
where $|\alpha|^2$ is the input power and $(G+g)^{-2}$ is the degree of squeezing at the dark port input. When $(G+g)^{-2}=1$, the above equation describes a conventional MZI with standard quantum limit (SQL) sensitivity. Following Eq.~(\ref{SQ-MZI1}), we {show how} a phase coded squeezed state evolves with loss in the left part of Fig.~\ref{fig:setup}(d). We note the noise is extremely sensitive to loss since its noise level is lower than that of a vacuum state.   

The SISNI method can help us to overcome the loss {sensitivity,} as shown in Fig.~\ref{fig:setup}(b). A nondegenerate optical parametric amplifier is used to generate twin beams $ \hat{c} $ and $ \hat{d} $, then the signal mode $ \hat{d} $ is injected into the {dark} input port of the MZI. When the MZI is locked at a dark fringe, the noise at the dark output will be still correlated to {the ``idler''} mode $\hat{e}$. Finally, mode $ \hat{h} $ and $ \hat{e} $ are injected into the second PA{, where destructive interference partially cancels} quantum noise at the output. Here the noise term $ \hat{b}_{\MZIsub} $ in Eq.~(\ref{MZI-inout}) is replaced by one of the twin beams, i.e. $\hat{d}_{\MZIsub}= (G_{1}\hat{a}+g_{1}\hat{b}^{\dagger})e^{\theta/2}$, 
where $\theta$ is the common-mode phase of the SU(1,1) interferometer paths. The relative phase is irrelevant to the signal (see Supplemental Material \cite{PRLSIRef}). The optimized SNR of SISNI is{
\begin{eqnarray}
\label{SISI}
\SNR_{\rm \acronym}&=&
 \frac{\eta_{s}G_{2}^{2}|\alpha|^2 \phisig^2}
 {L+ (\eta_{s}G_2^2 + \eta_{i}g_2^2)(G_1^2 +g_1^2)  -4\sqrt{\eta_{s}\eta_{i}}G_1 G_2 g_{1}g_{2}},
 \nonumber \\ 
\end{eqnarray} 
}
where $L=L_{e}+g_{2}^2(1-L_{e})L_{ii}+G_{2}^2(1-L_{e})L_{is} $ and $L_{\beta} $ with $ \beta\in\{is,ii,e\} $ indicate, respectively, internal loss of the signal mode [including loss in the SU(2) interferometer] and idler mode of the SU(1,1) interferometer, and  external loss after PA$_{2}$. $G_1=\sqrt{1+g_1^2}$ and $G_2=\sqrt{1+g_2^2}$ are the amplification gains of PA$_{1} $ and PA$_{2} $, respectively, where $1/(G_1+g_1)^2$ indicates the degree of two-mode squeezing generated by PA$_{1}$. The right part of Fig.\ref{fig:setup}(d) represents Eq.~(\ref{SISI}), which shows the same noise performance as SUI. We note that the SISNI improves the SNR relative to the conventional MZI with the same $|\alpha|^2$. 

To demonstrate the above-described quantum advantage at large signal strength, we implemented a \acronym~as shown in Fig.~\ref{fig:Noisespectral}(a).  PA$_{1}$ and PA$_{2}$ are implemented as FWM processes \cite{mccormick2008strong} in $^{85}$Rb, with amplification gains $G_1$ and $G_2$. A MZI, formed by two linear beam splitters and mirrors with piezoelectric transducers (PZTs), is nested in one arm of the SUI. The two input ports of PA$ _{1} $ are fed with vacuum to avoid excess noise from the FWM process. Laser light is injected into the bright input port of the MZI. Local oscillator (LO) beams are generated by FWM process to implement balanced homodyne detection.  We note that the interferometer becomes a simple MZI if the PAs' pump light is blocked. The relative phase $\phi$ of the two PA pumps is locked to minimum net amplification by a quantum noise locking technique \cite{mckenzie2005quantum}, and a coherent modulated locking technique is used to maintain the MZI at the dark fringe condition  \cite{du2018absolute}. {To lock the phase of the LO on the phase quadrature, we use a method described by Liu et al.  \cite{Liu:18}, in which the input coherent state is amplitude modulated and the envelope of modulations seen at the HD is fed back to the LO phase.} 

The output performance of \acronym~and MZI under the same operating conditions are shown in Fig.~\ref{fig:Noisespectral}. The black trace {in Fig.~\ref{fig:Noisespectral}(b)} is the output noise level of MZI at the dark fringe, which is also the vacuum noise level. The blue trace {in Fig.~\ref{fig:Noisespectral}(b)} shows the noise level of PA$_{2}$, \SI{6}{\decibel} above vacuum noise. To have an experimentally accessible measure of PA gain, we define the {quantum noise gain} (QNG) as $G_{q} \equiv \langle \delta^{2}\hat{X}(\theta) \rangle$ resulting from vacuum inputs. For an ideal PA, ${G_{q}=G^2+g^2}$. The red trace {in Fig.~\ref{fig:Noisespectral}(b)} shows noise reduction (minimum) and antireduction (maximum) by scanning the phase of twin beams with the MZI locked at the dark fringe. It shows \SI{2.4}{\decibel} of noise reduction below the noise level of PA$_{2} $, while the QNG of PA$_{1} $ is set as 4 dB. The inset {in Fig.~\ref{fig:Noisespectral}(c)} shows the power spectrum of MZI (black trace) and \acronym~(red trace) when a signal at \SI{1.7}{\mega\hertz} is introduced by modulating the PZT. As above, both the MZI and SUI are locked at their respective dark conditions.  The measured SNRs  \cite{kong2013cancellation} of  the MZI and \acronym~are \SI{4.8\pm0.2}{\decibel} and \SI{7.0\pm0.3}{\decibel}, respectively, which indicates a \SI{2.2\pm0.5}{\decibel} SNR enhancement.

 \begin{figure}[t]
	\centering
	\includegraphics[width=0.85\columnwidth]{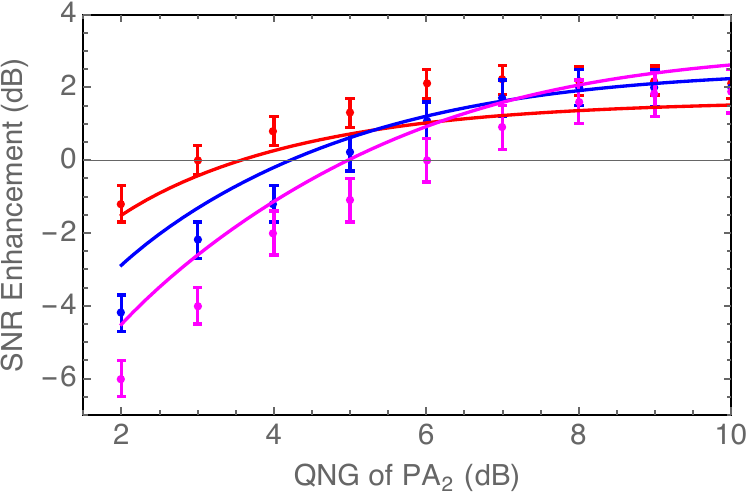}	
	\caption{Characterization of the trade-off between gain and FWM noise in the \boldacronym.  Horizontal axis shows QNG of PA$_2$, adjusted by control of the pump power.  Vertical axis shows the SNR advantage of the \acronym~over the MZI at equal coherent state input power, with positive values indicating advantage for the \acronym.   Points and error bars indicate mean and $\pm 1$ standard deviation from repeated measurements.  Red, blue and magenta data points and curves correspond to PA$_1$ QNG levels $\SI{4}{}$, $\SI{6}{}$ and $\SI{8}{\decibel}$, respectively.  Curves show fits based on {a minimal PA noise model \cite{PRLSIRef, OuPRA2012}.}   }
	\label{fig:Fitting}
\end{figure}
 
The above results were obtained for optimal gains $G_{q1} = \SI{4}{}$ and $G_{q2} = \SI{6}{\decibel}$. The existence of an optimum is an important characteristic of the SISNI with atomic FWM, not present in models with ideal PAs.  As shown in Fig.~\ref{fig:Fitting}, the SNR enhancement, measured as described above, increases with increasing PA$_2$ gain until saturation, and in the range studied decreases with increasing PA$_1$ gain. This behavior can be understood as the result of atomic dephasing: this introduces excess noise in the PA$_1$ output, which is further amplified by PA$_2$.    A detailed model including atomic dephasing and internal losses of 16\% and 10\% in the  signal and idler arms, respectively, agrees well with all experimental results (see Supplemental Material \cite{PRLSIRef}).

To verify that our scheme is able to maintain the SNR enhancement at  high photon flux, we measure the SNR of MZI and \acronym~while increasing the laser input power. As can be seen in Fig.~\ref{fig:Noisespectral}(d), both MZI and \acronym~performance are well fit by ${\rm SNR} = A |\alpha|^2$, which confirms their photon shot-noise limited performance. By the fit, the \acronym~SNR is \SI{2.2}{\decibel} above that of the MZI, comparable to that seen in other SUI experiments
\cite{RN18, RN19, RN20}. The advantage is observed up to nearly \SI{1}{\milli\watt}, which is hundreds of times higher than the \SI{}{\micro\watt}  levels of previous SUIs  \cite{hudelist2014quantum, manceau2017detection, linnemann2016quantum}. The interferometer, built from dielectric mirrors and beam splitters, could withstand much higher powers. In a SISNI the PAs need not support high powers.

%
%

The SISNI is tolerant of loss in the detection process, because the PA process boosts the signal above the vacuum noise level. PA processes, e.g. FWM, have been demonstrated from radio \cite{FedericiIEETPS1991} to XUV \cite{BencivengaN2015} wavelengths, whereas high-quantum-efficiency detectors exist for a much more limited range. Provided these PA processes can be quantum noise limited, the SISNI will enable sensitive, quantum-enhanced measurements in previously inaccessible spectral regions.  The SISNI, like the SUI, can use one wavelength to sense and another to detect, as in \cite{LemosN2014}, and can make simultaneous measurement of phase and amplitude, one at each mode \cite{RN17}. Possible use cases include reduced-damage probing of materials \cite{WolfgrammNP2013} and nano-imaging \cite{PooserPRL2020}.


In conclusion, we have experimentally demonstrated a new interferometer topology, the SU(2)-in-SU(1,1) nested interferometer. A detailed analysis shows that this topology combines the loss tolerance of SU(1,1) interferometry with the large signal strength of SU(2) interferometry.  Experimentally, we have demonstrated \SI{2.2}{\decibel} of signal-to-noise ratio improvement beyond the standard quantum limit, with optical power levels, and thus signal-to-noise ratios, beyond the reach of traditional SU(1,1) interferometry. 


%

\begin{acknowledgments}
W. Z. acknowledges support by the National Key Research and Development Program of China under Grant  No.~2016YFA0302001, the National Natural Science Foundation of China under Grants  No.~11654005, the Quantum Information Technology, Shanghai Municipal Science and Technology Major Project under Grant No. 2019SHZDZX01, and also additional support from the Shanghai talent program. W. D. acknowledges the support from the fellowship of China Postdoctoral Science Foundation under Grant  No.~2021M702147. J. K. acknowledges the support from NSFC through Grants  No.~12005049 and  No.~11935012. G. Z. B. acknowledges the support from the fellowship of China Postdoctoral Science Foundation under Grants  No.~2020TQ0193 and  No.~2021M702146. P. Y. Y. acknowledges the support from the fellowship of China Postdoctoral Science Foundation under Grant  No.~2021M702150. J. J. acknowledges support from the China Scholarship Council. C. H. Y. acknowledges the support by the National Natural Science Foundation of China under Grant  No.~11974111. J. F. C. acknowledges the Guangdong Provincial Key Laboratory under Grant  No.~2019B121203002 and  No.~2019ZT08X324. Z. Y. O. acknowledges the U.S. National Science Foundation under Grant  No.~1806425. M. W. M. acknowledges H2020 Future and Emerging Technologies Quantum Technologies Flagship projects MACQSIMAL (Grant Agreement  No.~820393) and QRANGE (Grant Agreement  No.~820405); H2020 Marie Sk?odowska-Curie Actions project ITN ZULF-NMR (Grant Agreement  No.~766402); Spanish Ministry of Science ``Severo Ochoa'' Center of Excellence CEX2019-000910-S, and project OCARINA [PGC2018-097056-B-I00 project funded by MCIN(Ministerio de Ciencia e Innovaci\'{o}n)/AEI /10.13039/501100011033/ FEDER ``A way to make Europe'']; Generalitat de Catalunya through the CERCA program; Ag\`{e}ncia de Gesti\'{o} d'Ajuts Universitaris i de Recerca Grant  No.~2017-SGR-1354; Secretaria d'Universitats i Recerca del Departament d'Empresa i Coneixement de la Generalitat de Catalunya, cofunded by the European Union Regional Development Fund within the ERDF Operational Program of Catalunya (project QuantumCat, ref. 001-P-001644); Fundaci\'{o} Privada Cellex; Fundaci\'{o} Mir-Puig; 17FUN03 USOQS, which has received funding from the EMPIR programme cofinanced by the Participating States and from the European Union's Horizon 2020 research and innovation program.

\end{acknowledgments}

%
%

\bibliographystyle{apsrev4-1no-url}
\bibliography{biblio210307}


\clearpage

\setcounter{page}{1}
\begin{widetext}
\noindent
{\Large Supplementary Information for \textit{\thetitle} }\\

{
\noindent\textbf{Definitions and notation.}  We will calculate the SNR of both squeezed-light-injected MZI  (\SQMZI)  and \acronym. The respective topologies and field mode names are shown in Figs.~\ref{fig:SQI} and \ref{fig:SU2Qutumstate}.  To avoid confusion with the \acronym~discussion that follows, we label MZI modes with the subscript $_\MZIsub$. We write the interferometer phase as $\varphi = \varphi_0 + \phisig$, where $\varphi_0$ is the set-point of the interferometer and $\phisig \ll \pi$ is the signal, i.e., a small phase excursion to be measured, assumed constant (but unknown) during the measurement. We write $\phisighat$ for the estimator of  $\phisig$, with value
\begin{equation}
\label{eq:PhiSigHatDef}
\phisighat \equiv \left( \cX-  \langle \cX \rangle_0  \right) \left(\frac{d{ \langle\cX\rangle}}{d \varphi} \right)^{-1}
\end{equation}
where $\cX$ is the measurement result, $\langle \cdot \rangle_0$ indicates an expectation taken with $\varphi = \varphi_0$, and derivatives with respect to $\varphi$ are understood to be taken at $\varphi_0$. The SNR is defined as 
\begin{equation}
\label{eq:SNRDef}
\SNR \equiv \frac{\phisig^2}{\langle \delta \phisighat^2 \rangle}  = {\frac{\phisig^2}{\langle \delta \cX^2 \rangle_0}
 \left| \frac{d{ \langle\cX\rangle}}{d \varphi}\right|^2}  
 = \frac{ \left(\langle \cX \rangle-  \langle \cX \rangle_0  \right)^2}{\langle \delta \cX^2 \rangle_0}
\end{equation}
where $\langle \delta \phisighat^2 \rangle$ is the variance, or equivalently the mean squared error (MSE), of the estimator and $\langle \delta \cX^2 \rangle_0 \equiv \langle \cX^2\rangle_0 -  \langle \cX  \rangle_0^2$ is the intrinsic variance of $\cX$ when $\varphi = \varphi_0$.  }

\noindent\textbf{Input-output relations, SQ-MZI.}   The quantum optical performance of each {strategy} can be analyzed by considering cascaded linear input-output relations, i.e., by a sequence of linear transformations on field operators. For the  \SQMZI, these are
\begin{eqnarray}
\hat{b}_{\MZIsub}&=&G\hat{a}_{\MZIsub}+g\hat{a}_{\MZIsub}^{\dag}\\
\hat{c}_{\MZIsub}&=&\sqrt{T}\hat{A}_{\MZIsub}-\sqrt{R}\hat{b}_{\MZIsub}\\
\hat{d}_{\MZIsub}&=&\sqrt{T}\hat{b}_{\MZIsub}+\sqrt{R}\hat{A}_{\MZIsub}\\
\hat{f}_{\MZIsub}&=&e^{i\varphi/2}\sqrt{1-L_{i}}\hat{c}_{\MZIsub}+\sqrt{L_{i}}\hat{C}_{\MZIsub}\\
\hat{g}_{\MZIsub}&=&e^{-i\varphi/2}\sqrt{1-L_{i}}\hat{d}_{\MZIsub}+\sqrt{L_{i}}\hat{B}_{\MZIsub}\\
\hat{i}_{\MZIsub}&=&\sqrt{T}\hat{g}_\MZIsub-\sqrt{R}\hat{f}_{\MZIsub}\\
\hat{k}_{\MZIsub}&=&\sqrt{1-L_{e}}\hat{i}_\MZIsub+\sqrt{L_{e}}\hat{D}_{\MZIsub},
\end{eqnarray} 
where $G=\sqrt{1+g^2}$ is the amplification gain, $L_{i}$ and $L_e$ indicate internal and external losses, respectively.  $\hat{B}_{\MZIsub}, \hat{C}_{\MZIsub}$ and $\hat{D}_{\MZIsub}$ describe vacuum modes introduced by the loss processes. Because $g$ and $G$ are positive real, the phase quadrature is squeezed by the PA.   The input  is vacuum for mode {$\hat{a}_\MZIsub$} and a coherent state $|\alpha\rangle$, $\alpha$ positive real, for $\hat{A}_{\MZIsub}$. For $T=R=1/2$, the case of interest, the output $\hat{k}_{\MZIsub}$ simplifies to 
\begin{eqnarray}
\hat{k}_{\MZIsub}&=&\hat{N}_{\MZIsub}+\sqrt{\eta}\left[(G\hat{a}_{\MZIsub}+g\hat{a}_{\MZIsub}^{\dagger})\cos\frac{\varphi}{2}-i\hat{A}_{\MZIsub}\sin\frac{\varphi}{2}\right],
\end{eqnarray}
where $\hat{N}_{\MZIsub}=\sqrt{L_{e}}\hat{D}_{\MZIsub}+\sqrt{L_{i}(1-L_{e})/2}(\hat{B}_{\MZIsub}-\hat{C}_{\MZIsub}) $ and $\eta=(1-L_{i})(1-L_{e})$.

The signal is given by the phase quadrature $\cX =  X_{2,\MZIsub} \equiv i(\hat{k}_{\MZIsub}^{\dag}-\hat{k}_{\MZIsub})$. The  signal is 
\begin{eqnarray}
\langle {\cal X} \rangle =-2 \alpha \sqrt{\eta} \sin \frac{\varphi}{2}.
\end{eqnarray} 

The signal slope is 
\begin{eqnarray}
\frac{d\left\langle {\cal X} \right\rangle }{d\varphi}=-\alpha \sqrt{\eta}\cos\frac{\varphi}{2}.
\end{eqnarray} 

The noise variance is 
\begin{eqnarray}
\langle \delta \cX^2 \rangle_0   & = & 
\langle\delta^2X_{2,\MZIsub}\rangle_0 \nonumber \\ 
&=&\langle X_{2,\MZIsub}^2\rangle_0-\langle X_{2,\MZIsub}\rangle_0^2\nonumber\\&=&1+\eta\left[ (G-g)^2-1\right] \cos ^2\frac{\varphi_0}{2}
\end{eqnarray} 
{
The MSE for the phase estimate is
\begin{eqnarray}
\langle \phisighat^2 \rangle_0 &=& \langle \delta \cX^2 \rangle_0   \left
|\frac{d{ \langle\cX\rangle}}{d \varphi} \right|_{\varphi = \varphi_0}^{-2}
\nonumber \\ &= &  \frac{ \eta[(G-g)^2-1] + \sec^2\frac{\varphi_0}{2}}{\eta|\alpha|^2},
\end{eqnarray} 
Which is minimized for $\varphi_0=0$, which defines the optimal operating point. The SNR is \begin{eqnarray}
\SNR_{\SQMZI} &=&\frac{\eta\left| \alpha\right| ^2 \phisig^2 \cos^2 \frac{\varphi_0}{2}}{1+\eta\left[ (G-g)^2-1\right] \cos^2\frac{\varphi_0}{2} },
\end{eqnarray} 
or at the optimal operating point:
\begin{eqnarray}
\SNR_{\SQMZI} &=&\frac{\eta\left| \alpha\right| ^2 \phisig^2  }{1+\eta\left[ (G-g)^2-1\right]}.
\end{eqnarray} 
}

\begin{figure}[t]
	\centering
	\includegraphics[width=12cm]{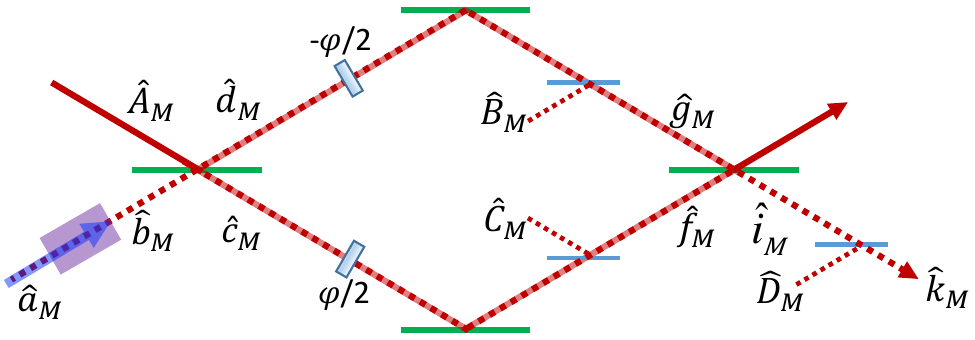}	
	\caption{Squeezed light injected MZI. }
	\label{fig:SQI}
\end{figure}

\begin{figure}[t]
	\centering
	\includegraphics[width=12cm]{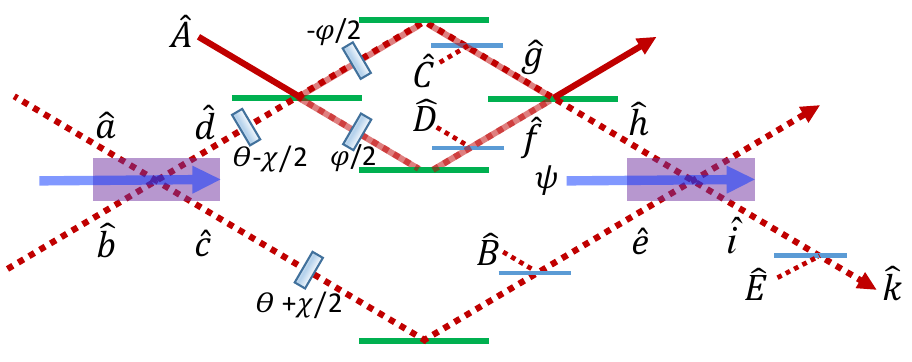}	
	\caption{SU2-in-SU(1,1) nested interferometer. }
	\label{fig:SU2Qutumstate}
\end{figure}

\noindent\textbf{Input-output relations, SISNI.} 
For the \acronym, the linear transformations are 
\begin{eqnarray}
\label{eq:InOutC}
\hat{c}&=&e^{i(\theta+\chi/2)}(G_{1}\hat{a}+g_{1}\hat{b}^{\dag})\\
\label{eq:InOutD}
\hat{d}&=&e^{i(\theta-\chi/2)}(G_{1}\hat{b}+g_{1}\hat{a}^{\dag})\\
\hat{e}&=&\sqrt{1-L_{ii}}\hat{c}+\sqrt{L_{ii}}\hat{B}\\
\hat{f}&=&e^{i\varphi/2}\sqrt{1-L_{is}}(\sqrt{T}\hat{A}-\sqrt{R}\hat{d})+\sqrt{L_{is}}\hat{C}\\
\hat{g}&=&e^{-i\varphi/2}\sqrt{1-L_{is}}(\sqrt{T}\hat{d}+\sqrt{R}\hat{A})+\sqrt{L_{is}}\hat{D}\\
\hat{h}&=&\sqrt{T}\hat{g}-\sqrt{R}\hat{f}\\
\label{eq:InOutG}
\hat{i}&=&e^{i\psi}G_{2}\hat{h}+e^{-i\psi}g_{2}\hat{e}^{\dag}\\
\label{eq:InOutH}
\hat{k}&=&\sqrt{1-L_{e}}\hat{i}+\sqrt{L_{e}}\hat{E},
\end{eqnarray}
where $G_1=\sqrt{1+g_1^2}$ and $G_2=\sqrt{1+g_2^2}$ are the amplification gains of PA$ _{1} $ and PA$ _{2} $,  $L_{\beta} $ with $ \beta\in\{is,ii,e\} $ indicates internal loss affecting the signal mode ($\hat{d}$) and idler mode ($\hat{c}$) and external loss after PA$_{2}$. $\hat{B}, \hat{C}, \hat{D}, \hat{E}$ and $\hat{F}$ are vacuum modes introduced by the loss channels, $\varphi$ is the sensing phase, $ \psi $ is the pump phase at the second PA and $\theta$ is the common-mode phase of signal and idler.

By a sequence of substitutions, the output mode $\hat{k}$ is found to be: 
{
\begin{eqnarray}
\hat{k}&=&N_{i}-i\sqrt{\eta_{s}}G_{2}\sin\frac{\varphi}{2}e^{i  \psi}\hat{A}
+ 
(\sqrt{\eta_{i}}e^{-i(\theta + \psi)}G_{1}g_{2}+\sqrt{\eta_{s}}e^{i(\theta + \psi)}g_{1}G_{2}\cos\frac{\varphi}{2})e^{-i \chi/2}\hat{a}^{\dagger}
\nonumber\\& &+
(\sqrt{\eta_{i}}e^{-i (\theta + \psi)}g_{1}g_{2}+\sqrt{\eta_{s}}e^{i (\theta + \psi) }G_{1}G_{2}\cos\frac{\varphi}{2})e^{-i \chi/2}\hat{b}, \hspace{6mm} 
\end{eqnarray}
where $\hat{N}_{i}=\sqrt{L_{e}}\hat{E}+e^{-i\psi}g_{2}\sqrt{L_{ii}(1-L_{e})}\hat{B}^\dagger-e^{i\psi}G_{2}(\sqrt{L_{is}(1-L_{e})/2}(\hat{C}-\hat{D}) $, and $ \eta_{s}=(1-L_{is})(1-L_{e})$, $ \eta_{i}=(1-L_{ii})(1-L_{e})$.  Without loss of generality, we absorb the factors $\exp[-i \chi/2]$ into $\hat{a}^\dagger$ and $\hat{b}$. 

The signal is given by the phase quadrature $\cX =  X_{2} \equiv i(\hat{k}^{\dag}-\hat{k})$. The mean signal is 
\begin{eqnarray}
\langle \cX \rangle  =- 2 G_{2}\sqrt{\eta_{s}}\alpha\sin\frac{\varphi}{2}\cos\psi,
\end{eqnarray}
and the signal slope is 
\begin{eqnarray}
\frac{d	\langle \cX \rangle}{d\varphi} =-G_{2}\sqrt{\eta_{s}}\alpha\cos\frac{\varphi}{2}\cos\psi.
\end{eqnarray}
The variance is 
\begin{eqnarray}
\langle\delta^2 \cX\rangle_0&=&\langle X_{2}^2\rangle_0-\langle  X_{2}\rangle_0^2\nonumber\\&=&L_{e}+(1-L_{e})(G_{2}^{2}+g_{2}^{2})+(G_{1}^{2}+g_{1}^{2}-1)(G_{2}^2\eta_{s}\cos^2\frac{\varphi_0}{2}+g_{2}^{2}\eta_{i})+4G_{1}g_{1}G_{2}g_{2}\sqrt{\eta_{s}\eta_{i}}\cos\frac{\varphi_0}{2}\cos2(\theta+\psi). \hspace{6mm}
\end{eqnarray}

The MSE of the estimator is
\begin{eqnarray}
\langle \delta \hat{\varphi}^2 \rangle_0 &=&\frac{L_{e}+(1-L_{e})(G_{2}^{2}+g_{2}^{2})+(G_{1}^{2}+g_{1}^{2}-1)(G_{2}^2\eta_{s}\cos^2\frac{\varphi_0}{2}+g_{2}^{2}\eta_{i})+4G_{1}g_{1}G_{2}g_{2}\sqrt{\eta_{s}\eta_{i}}\cos\frac{\varphi_0}{2}\cos 2(\theta+\psi)}{G_2^2 \eta_s |\alpha|^2 \cos^2\frac{\varphi_0}{2}\cos^2\psi}. \hspace{6mm}
\end{eqnarray} 
This MSE is minimized for  $\varphi_0 = \psi = 0$ and $\theta = \pi/2$, which defines the optimal operating point.  At this point, $\langle \cX \rangle_0 = 0$, which is the dark fringe condition, and the SNR is  
{
\begin{eqnarray}
\SNR_{\rm \acronym}&=&
 \frac{\eta_{s}G_{2}^{2}|\alpha|^2 \phisig^2}
 {L+ (\eta_{s}G_2^2 + \eta_{i}g_2^2)(G_1^2 +g_1^2)  -4\sqrt{\eta_{s}\eta_{i}}G_1 G_2 g_{1}g_{2}}
 \nonumber \\ 
\end{eqnarray} 
}
where $ L=L_{e}+g_{2}^2(1-L_{e})L_{ii}+G_{2}^2(1-L_{e})L_{is} $.

In the simplest case of no loss and $G_1 = G_2 =  G$, the SNR of the SQ-MZI and  SISNI reduce to 
\begin{eqnarray}
	\SNR_{\SQMZI}&=&\frac{\left| \alpha\right| ^{2}\phisig^{2}}{(G-g)^{2}}
\end{eqnarray} 
and
\begin{eqnarray}
	\SNR_{\rm \acronym}&=&G^{2}\left| \alpha\right| ^{2}\phisig   ^{2},
\end{eqnarray} 
respectively.  For large gain $G$ and $g = \sqrt{G^2-1}$, $G^2/(G-g)^{-2} \rightarrow 1/4$. The difference in the sensitivity at equal gain reflects the fact that in the SQ-MZI, all photons produced in the PA process experience the sensing phase shift $\varphi$, whereas in the \acronym, only one photon of each pair experiences this phase shift.  In the absence of losses, a \acronym~with gain $G_{\acronym}$ and a \SQMZI~with gain $G_{\SQMZI}$ will have equal SNR when 
\begin{equation}
\frac{\SNR_{\acronym}}{\SNR_{\SQMZI}} = \frac{G^2_{\acronym}}{(G_{\SQMZI}-g_{\SQMZI})^{-2}} \approx \frac{G^2_{\acronym}}{4 G^2_{\SQMZI}} = 1,
\end{equation}
or $G_{\acronym} = 2 G_{\SQMZI}$.  In practice it is possible to have arbitrarily large gain with either one-mode or  two-mode squeezing by, e.g.,  enclosing the parametric gain material in a resonator to produce a parametric oscillator.  Because of this, the ultimate performance of an enhancement technique will not be limited by the available gain, but rather by the relationship between gain and deleterious effects created in the amplification process, e.g.  loss and excess noise. }

~ 

\noindent\textbf{Parametric amplifier noise model.}  
\label{sec:Modeling}
To understand the gain-advantage relationship described in the previous section and seen in \autoref{fig:Fitting}, we must modify the noise model of the PAs, which until this point have been described as ideal two-mode squeezers.  An accurate and detailed model of noise in FWM processes is beyond the scope of this manuscript, but we can obtain qualitative insights by adapting a simple physical model described previously \cite{OuPRA2012}. In this model{,} PA$_1$ has the input-output relations 
\begin{eqnarray}
\hat{c}&=&\bar{G}_{1}\hat{b}+\bar{g}_{1}\hat{a}^{\dag} + \bar{G}_{1}' \hat{b}_0 +   \bar{g}_{1}' \hat{a}_0^\dagger \\
\hat{d}&=&\bar{G}_{1}\hat{a}+\bar{g}_{1}\hat{b}^{\dag} + \bar{G}_{1}' \hat{a}_0 +   \bar{g}_{1}' \hat{b}_0^\dagger,
\end{eqnarray}
where $\hat{a}_0$ and $\hat{b}_0$ are auxiliary modes.  Here we take these to be in thermal states with quadrature variances $\langle \Delta X^2 {\rangle} = \langle \Delta P^2 \rangle \equiv \epsilon^2 \ge 1$.  The input-output relations for PA$_2$ are the same as above with the substitutions $\hat{a}\rightarrow \hat{e}$, $\hat{b}\rightarrow \hat{f}$, $\hat{c}\rightarrow \hat{g}$, $\hat{d}\rightarrow \hat{h}$, and $1\rightarrow 2$.  The coupling factors are (subscripts $i = 1,2$ indicating which PA are henceforth omitted): $\bar{G}  =  [(1 - \rho^2)/4 + |\kappa|^2]/M$, $\bar{g} = \kappa/M$,  $\bar{G}'  =  \sqrt{\rho}(1 + \rho)/(2M)$  and $\bar{g}' = \kappa \sqrt{\rho}/M$, where $M  =  (1 + \rho)^2/4 - |\kappa|^2$. Loss and gain are parametrized by  $\rho$  and $\kappa$, respectively.  This model originates in the description of lossy, cavity-based two-mode parametric amplifiers, and for this reason is not expected to match quantitatively the single-pass FWM we use here. Nonetheless, it has qualitative features appropriate to the FWM scenario, e.g. the noise introduced by loss is amplified by the gain process.  We note that for given $\epsilon^2$ and $\rho$,  $\kappa$ can be found from the QNG. 

To compare against the measured SNR enhancement, we use the measured loss values $L_{is}=0.16 $, $ L_{ii}=0.1 $, $ L_{e}=0.15$, and the above input-output relations in place of Eqs.~(\ref{eq:InOutC}),(\ref{eq:InOutD}) for PA$_1$ and Eqs.~ (\ref{eq:InOutG}), (\ref{eq:InOutG})  for PA$_2$.  By the same mathematical machinery described above we compute $\SNR_{\rm \acronym}$ and the SNR advantage relative to the SQL.  Fitting to the measured data we find best fit parameters $\rho_1/\gamma_1 =\SI{5e-4}{}$,  $\rho_2/\gamma_2 =\SI{4e-4}{}$, $\epsilon_1^2 =2$, $\epsilon_2^2 =208$.   We observe that the model reproduces the qualitative features of the data, as seen in Fig.~\ref{fig:Fitting}, including the observed improved performance at lower PA$_1$ gain levels. 

\bibliographystyleSI{apsrev4-1no-url}

\end{widetext}
\end{document}